\documentclass[aps,prc,reprint,nofootinbib]{revtex4-1}
\usepackage{graphicx}   %       for graphics
\usepackage{latexsym}   %       for special symbols
\begin{document}
%==============================================================================
\title{Influence of the hadronic phase on observables in ultrarelativistic heavy ion collisions}

\author{J. Steinheimer$^1$, J. Aichelin$^{1,3}$, M. Bleicher$^{1,2,5}$, and H. St\"ocker$^{1,2,4}$}

\affiliation{$^1$ Frankfurt Institute for Advanced Studies, Ruth-Moufang-Str. 1, 60438 Frankfurt am Main, Germany}
\affiliation{$^2$ Institut f\"ur Theoretische Physik, Goethe Universit\"at Frankfurt, Max-von-Laue-Strasse 1, D-60438 Frankfurt am Main, Germany}
\affiliation{$^3$ SUBATECH, UMR 6457, Universit\'{e} de Nantes, Ecole des Mines de Nantes, IN2P3/CNRS. 4 rue Alfred Kastler, 44307 Nantes cedex 3, France}
\affiliation{$^4$ GSI Helmholtzzentrum f\"ur Schwerionenforschung GmbH, D-64291 Darmstadt, Germany}
\affiliation{$^5$ John von Neumann Institut f\"ur Computing, FZ J\"ulich, 52425 J\"ulich, Germany}
%\date{February 28, 2014}

\begin{abstract}
The hadronic phase in ultrarelativistic nuclear collisions has a large influence on final state observables like multiplicity, flow and $p_t$ spectra, as studied in the UrQMD approach. In this model one assumes that a non-equilibrium decoupling phase follows a fluid dynamical description of the high density phase. Hadrons are produced assuming local thermal equilibrium 
and dynamically decouple during the hadronic rescattering until the particles are registered in the detectors. This rescattering of hadrons modifies every hadronic bulk observable. The apparent multiplicity of resonances is suppressed as compared to a chemical equilibrium freeze-out model, because the decay products rescatter. Therefore the resonances, which decay in the early hadronic phase, cannot be identified anymore by the invariant mass method. Stable and unstable particles change their momentum distribution by more than 30$\%$ through rescattering and their multiplicity is modified by resonance production and annihilation on a similar magnitude. These findings show that it is all but trivial to conclude from the final state observables on the properties of the system at an earlier time where it may have been in local equilibrium.   
\end{abstract}

\maketitle

\section{Introduction}
An important outcome of the ultrarelativistic heavy ion experiments at the Relativistic Heavy Ion Collider (RHIC) and the Large Hadron Collider (LHC) is the assertion that a novel phase of matter, the so called quark gluon plasma (QGP) is created for a short time in these reactions \cite{Gyulassy:2004zy,Adams:2005dq,Back:2004je,Arsene:2004fa,Adcox:2004mh}. Such a state has been theoretically predicted in lattice gauge calculations (lQCD) which solve the equations of Quantum Chromo Dynamics (QCD) on a lattice under the condition that the system is in thermal and chemical equilibrium.

Present and future \cite{Hohne:2005qm} experiments aim at a better understanding of the properties of the QGP (see e.g. \cite{Friman:2011zz,nica}). This includes the determination of the yet unknown transport properties of QCD matter as well as the study of its phase structure at large net baryon densities. Both can presently not be calculated in lQCD calculations. This involves the study of hadron multiplicities but also fluctuations of conserved quantities in a given rapidity interval which can be compared with lQCD results\\

The latter studies are faced with the problem that lQCD simulations are limited to systems in thermal equilibrium whereas in heavy ion reactions we are confronted with rapidly expanding systems which may or may not hadronize at (or near) the crossover line (chemical freeze out) and continue to expand until the interactions cease. This latter phase is called hadronic phase and is characterized by hadronic interactions which govern the time period when the system drops out of equilibrium. If one wants to compare the measured observables with the results of lQCD at an earlier times where the system may be in local equilibrium, one has to assess how the hadronic phase has modified the observables. As we will see, these modifications are not negligible. For the fluctuations of conserved charges we have shown this already in a previous publication \cite{Steinheimer:2016cir}.   

To describe the expansion of the QGP and in particular the hadronic phase sophisticated model simulations are required. The current state-of-the-art of such models are so called hybrid models, where an initial state inferred from perturbative QCD calculations is used as input for a (viscous) fluid dynamical simulation \cite{Gale:2013da,Werner:2010aa,Shen:2014vra,Petersen:2008dd}. In the fluid dynamic approach the thermal properties of the matter under investigation, can be easily introduced, e.g. via the equation of state and the transport parameters. An important disadvantage of the fluid dynamical approach is that it is only valid for the description of systems at or very close to local thermal equilibrium. In general this condition seems to be fulfilled in the early phase of the system evolution, but eventually the densities of the rapidly expanding systems drop to a point where the mean free path becomes much larger than the system size. This usually occurs well after the expanding system has hadronized and can be described as a gas of interacting hadrons. The evolution of such a system can be described with the relativistic Boltzmann equation\\

\begin{equation}
\label{boltzmann}
p^\mu \cdot \partial_\mu f_i(x^\nu, p^\nu) = \mathcal{C}_i \quad .
\end{equation}

Here the change of the single particle distributions $f_i(x^\nu, p^\nu)$ (left hand side) is caused by the collision term $\mathcal{C}_i$ (right hand side).
Using a transport model (based on the Boltzman equation) as an afterburner which follows the fluid dynamical evolution has become the standard method of treating the interactions during the hadronic phase of nuclear collisions \cite{Bass:2000ib,Teaney:2000cw,Hirano:2005xf,Nonaka:2010zza,Hirano:2010jg,Gale:2013da,Ryu:2012at,Shen:2014vra,Petersen:2008dd}.
Recently several studies have pointed out the importance of the hadronic rescattering, during the freeze-out phase of a nuclear reaction, on the description of observables like particle numbers
and spectra, resonance production, momentum anisotropies and particle number fluctuations \cite{Knospe:2015nva,Steinheimer:2012rd,Steinheimer:2012era,Ilner:2016xqr}. \\

In this paper we will present a detailed analysis of the microscopic processes, occurring during the hadronic phase,
as described by the microscopic transport model UrQMD. The goal is to give a comprehensive summary of
the hadronic interactions during the freeze-out and how and why they affect the final state observables.
Furthermore we will give specific examples of how the hadronic phase can be experimentally verified, 
e.g. through hadronic resonance measurements.

\begin{figure}[t]	%       -----------------------------------------
\includegraphics[width=0.5\textwidth]{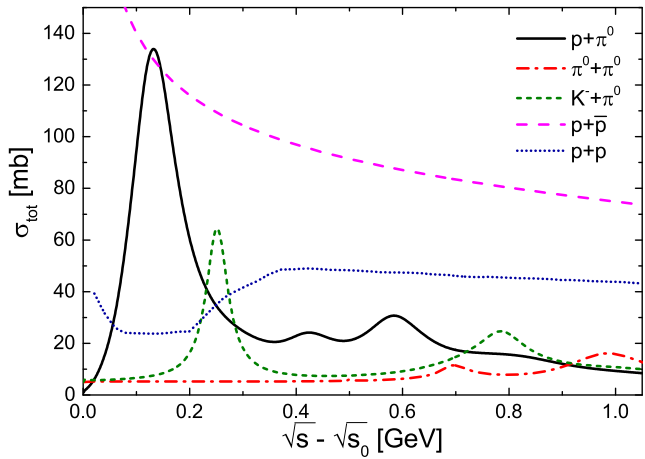}
\caption{[Color online] Different total (elastic + inelastic) hadronic cross sections as implemented in UrQMD. We show the cross sections as function of the relative energy, i.e. we subtract the rest masses $\sqrt{s_0}$ of both hadrons from the total invariant mass.
}\label{f0}
\end{figure}		%       -----------------------------------------  

\section{UrQMD} 
To study the dynamics of the hadronic phase of a heavy ion collision we will use the UrQMD hybrid model \cite{Petersen:2008dd}. This model combines a 3+1D ideal fluid dynamical simulation of the hot and dense phase of the collision (QGP phase) with a state-of-the-art transport description of the non-equilibrium hadronic decoupling phase. 
As the system at some point enters a dilute state, and is rapidly expanding, one expects that it quickly leaves a state of local thermal and chemical equilibrium. Therefore the fluid dynamical description is not valid anymore and a non-equilibrium transport description for the hadronic phase is necessary.
In the present investigation we use the Cooper-Frye equation,
\begin{equation}
\label{cooper_frye}
E \frac{dN}{d^3p}=\int_\sigma f(x,p) p^\mu d\sigma_\mu \,
\end{equation}
to transform the fluid dynamical fields to discrete hadrons on an iso-energy density hypersurface  \cite{Huovinen:2012is} $e_{CF}\approx$~350 MeV/fm$^3$.
Here $f(x,p)$ corresponds to the grand canonical Bose- or Fermi- distribution functions, depending essentially on the local temperature $T(x)$ and chemical potentials $\mu(x)$. To obtain a finite number of particles the integral is randomly sampled, under the constraint of baryon number, strangeness and net-charge conservation.
After the hadrons are "created" on the hypersurface they are able to interact in the transport part of the model description. 

\begin{figure}[t]	%       -----------------------------------------
\includegraphics[width=0.5\textwidth]{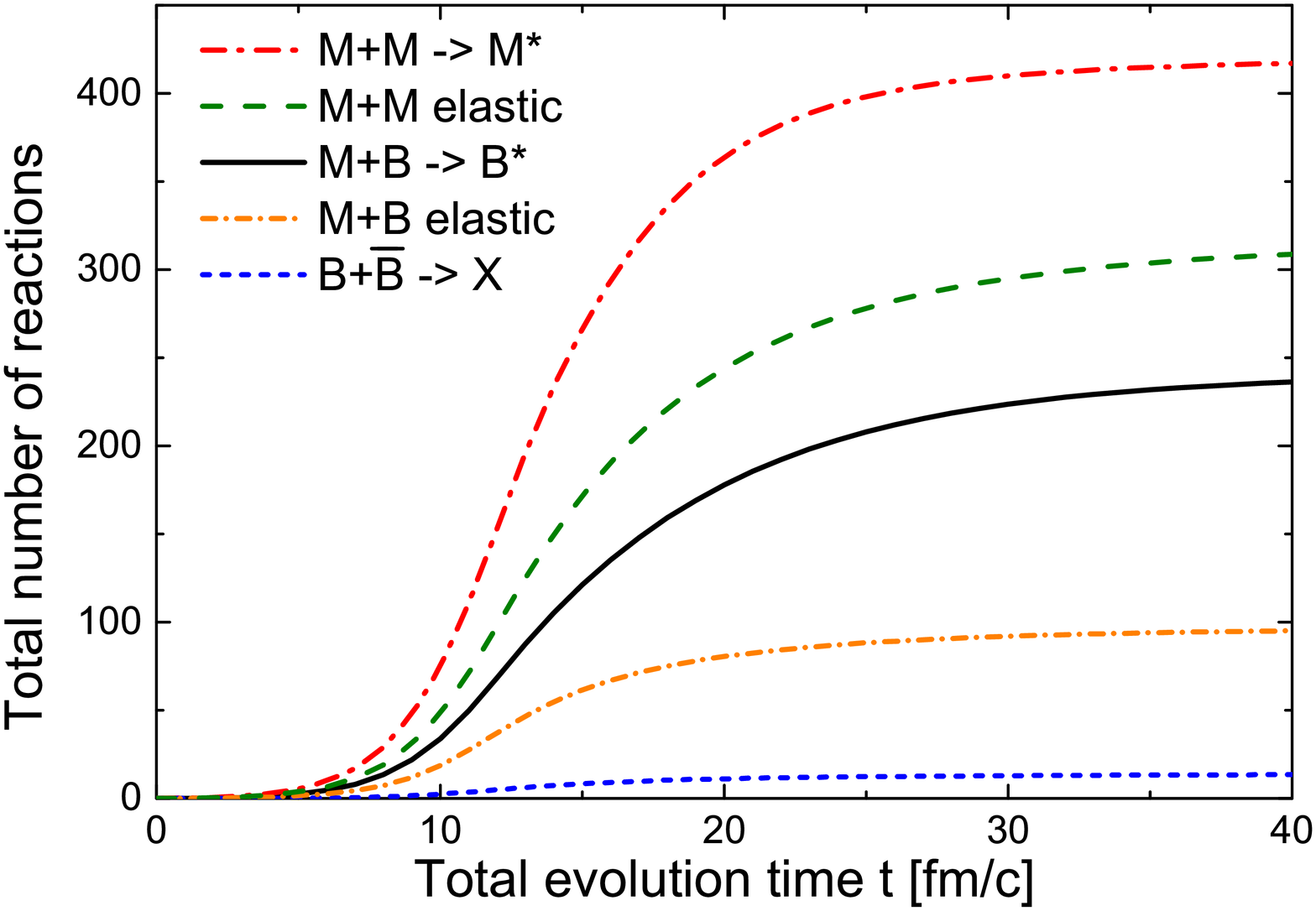}
\caption{[Color online] Total number of different hadronic interactions occurring until the time $t$, during the hadronic phase of central collisions of Au+Au
at $\sqrt{s_{NN}}= 200 $ GeV. 
}\label{f1}
\end{figure}		%       ----------------------------------------- 

The non-equilibrium transport part of the UrQMD model is based on $2\rightarrow n$ hadronic scattering, according to specific reaction cross sections, which serve as the main input to the model. The model includes nearly 60 different baryonic species with their anti-particles as well as about 40 mesonic states \cite{Bass:1998ca,Bleicher:1999xi}. The possible interactions between these hadrons include elastic scattering, resonance excitations, string excitations as well as strangeness exchange reactions. The cross sections are taken, when available, from the particle data group compilation \cite{Olive:2016xmw}. Unknown cross sections are mainly estimated using the additive quark model or taken from model calculations. To present an example of the cross sections implemented in the UrQMD model we show the total cross sections for several processes in Fig. \ref{f0}. As one can see there are several processes with very large cross sections, mainly involving protons. In particular the proton+anti-proton annihilation cross section  is large. To draw conclusions on the importance of these cross sections on final state observables one also requires knowledge on the phase space distributions and densities of the involved hadrons. The probabilities of certain reactions do not only depend on the magnitude of the cross section but also on the phase space densities of the involved hadrons. This makes the microscopic description of the hadronic phase so important.   

\section{Results}

We start out by giving estimates on the expected number of hadronic processes during and the duration of the hadronic phase. We will investigate the dynamics of the hadronic interactions in the final phase of collisions of Au+Au at beam energies of $\sqrt{s_{NN}}=200$ GeV,
focusing on the mid rapidity region $-0.5 < y < 0.5$ of most central ($b < 3.4$ fm) collisions where experimental data are available. For the following results we have generated 10000 events for each centrality bin presented.

\begin{figure}[t]	%       -----------------------------------------
\includegraphics[width=0.5\textwidth]{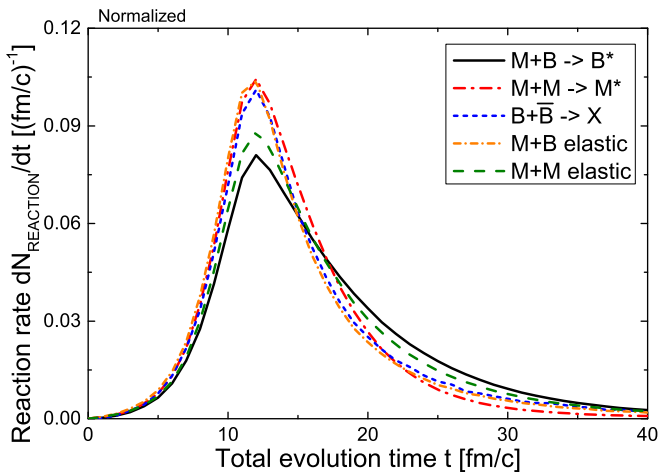}
\caption{[Color online] Reaction rates of different hadronic interactions during the hadronic phase of central collisions of Au+Au
at $\sqrt{s_{NN}}= 200 $ GeV, as function of the total system evolution time. The rates are normalized to the total number of the specific collisions in order to make different time dependencies visible.
}\label{f2}
\end{figure}		%       -----------------------------------------    

The total number of hadronic scatterings, up to the time $t$, defined as the time since the first initial collisions have taken place, is shown in figure \ref{f1}. Here we compare different possible hadronic interactions, meson+meson resonance excitation and elastic scattering, meson+baryon resonance excitation and elastic scattering 
as well as baryon+antibaryon annihilation.
For the collision energy under investigation the final rescattering is dominated by meson+meson interactions, due to the large number of mesons created 
at hadronization and only very few annihilation processes take place. However, since there are only few baryons created at hadronization the relative importance of these annihilations may still be significant, as we will see later. The number of scatterings changes most drastically in the time between hadronization, around 10 fm/c, and a time of roughly 20 fm/c, after which the number of hadronic scatterings does essentially not change anymore. This means that the hadronic rescattering mainly takes place in the 10 fm/c after hadronization.
To interpret these collision numbers one should note that roughly 100 baryons and anti-baryons are produced in the mid-rapidity region of a heavy ion collision. Thus the average number of resonance excitations per baryon is at least 2, or, in other words, every proton or neutron finds itself, on average, twice in an excited state during the expansion.

This result can also be seen in figure \ref{f2} which shows the normalized reaction rate for the different scattering processes displayed in figure \ref{f1}.
The scattering rate is normalized by the total number of scatterings to allow for a comparison of the shapes of the time dependence. We observe that the time evolution of the scattering rate for all processes is quite similar, or, in other words, the decrease of the density dominates over the size of the reaction cross section. At the end of the expansion the large $\Delta$ baryon excitation cross section leads only to a slightly larger reaction rate for M+B inelastic collisions.

\begin{figure}[t]	%       -----------------------------------------
\includegraphics[width=0.5\textwidth]{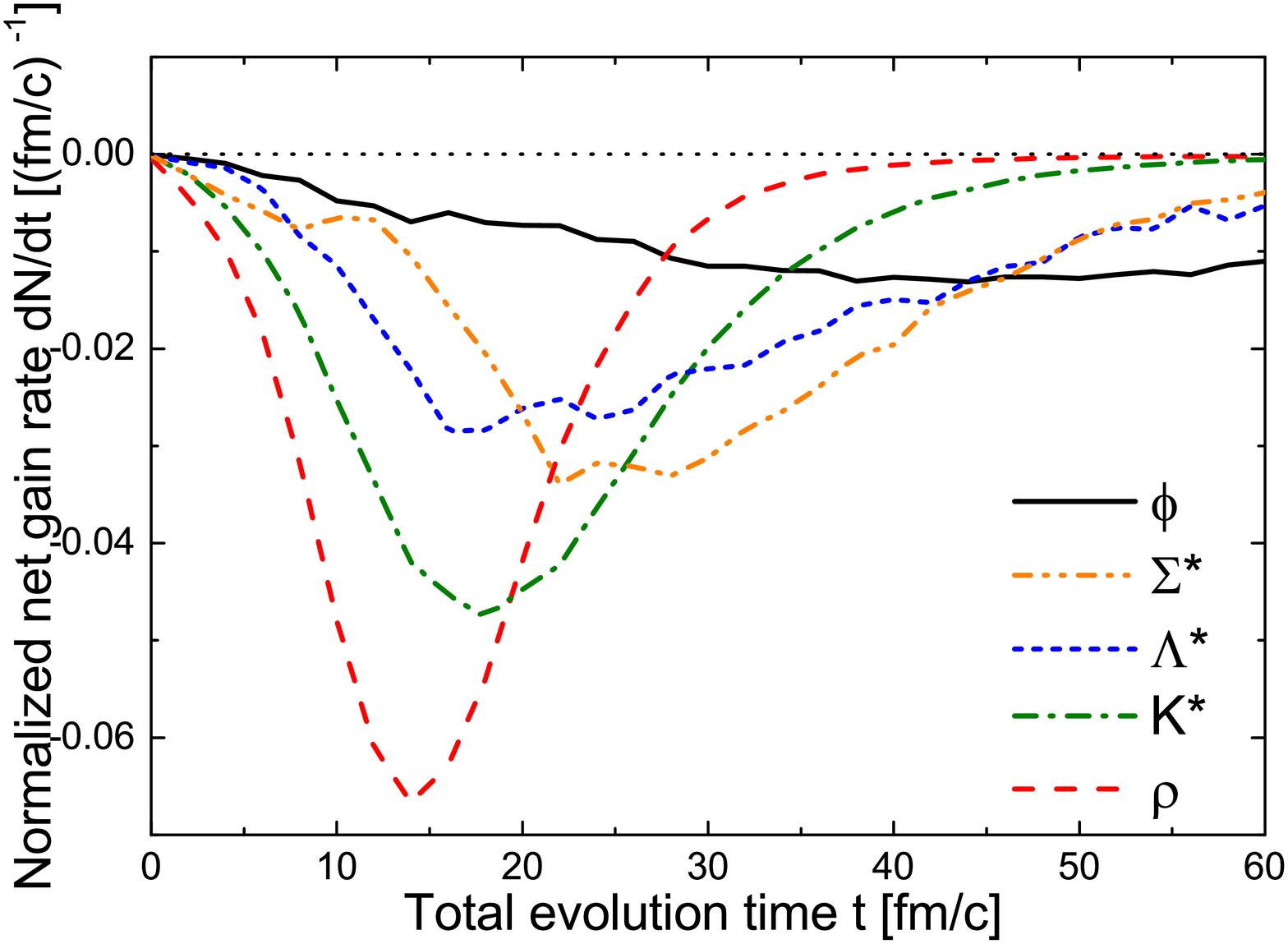}
\caption{[Color online] Net gain rates of different hadronic resonances as function of time and normalized to the initial yield at hadronization (Cooper-Frye). The results are for central collisions of Au+Au
at $\sqrt{s_{NN}}= 200 $ GeV.
}\label{f3}
\end{figure}		%       -----------------------------------------   

\subsection{Resonance dynamics}
If we want to verify experimentally the existence and duration of the hadronic phase, the study of the production and absorption rates of hadronic resonances in nuclear collisions is a useful tool \cite{Bleicher:2002dm}. Resonances which decay early during the hadronic phase may not be identified experimentally because the decay products rescatter. Therefore the number of identified resonances reflects the hadronic interaction during the hadronic phase.
There are several hadronic resonances which can be and have been experimentally reconstructed and may carry information of the hadronic phase. The most promising states, which will be studied in this work, are summarized in table \ref{t1}. 

\begin{figure}[t]	%       -----------------------------------------
\includegraphics[width=0.5\textwidth]{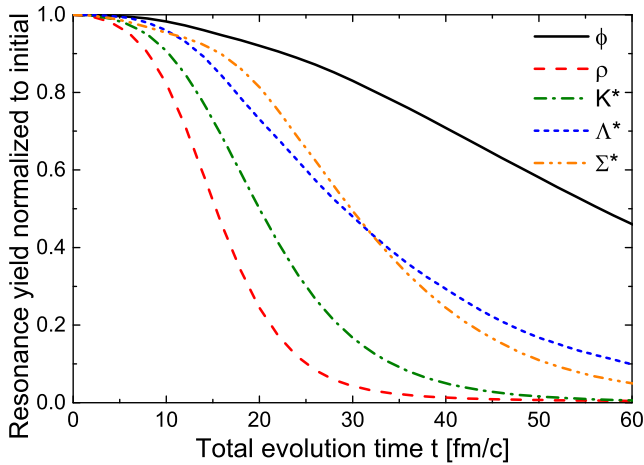}
\caption{[Color online] Yields of different hadronic resonances as function of time and normalized to the initial yield at hadronization (Cooper-Frye). One clearly sees the lifetime hierarchy. The results are for central collisions of Au+Au
at $\sqrt{s_{NN}}= 200 $ GeV.
}\label{f4}
\end{figure}		%       ----------------------------------------- 

\begin{table}[b]
\centering
\caption{Resonances and their decay channels}
\label{table2}
\begin{ruledtabular}
\begin{tabular}{llccc}
Resonance & decay channel & branching ratio & lifetime (fm/$c$) \\
\hline
$\rho(770)^{0}$ & $\pi^{+}$ + $\pi^{-}$  & 1 & 1.335 \\
$K^{*}(892)^{0}$ & $\pi^{-}$ + $K^{+}$ & 0.67 & 4.16 \\
$\phi(1020)$ & $K^{+}$ + $K^{-}$ & 0.489 & 46.26 \\
$\Sigma(1385)^{+}$ & $\pi^{+}$ + $\Lambda$ & 0.870 & 5.48 \\
$\Sigma(1385)^{-}$ & $\pi^{-}$ + $\Lambda$ & 0.870 & 5.01 \\
$\Lambda(1520)$ & $K^{-}$ + $p$ & 0.225 & 12.54\\
$\Xi(1530)^{0}$ & $\pi^{+}$ + $\Xi^{-}$ & 0.67 & 22 \\
\end{tabular}
\end{ruledtabular}
\label{t1}
\end{table} 

It is obvious that the listed hadronic resonances have widely varying properties and in particular different lifetimes.
For example, the $\phi$ meson has a more than 10 times larger lifetime than the $\rho$ meson which should be reflected in the 
dynamics of these resonances during the hadronic phase. To see that this is indeed true we show, in figure \ref{f3}, the net resonance
gain rate as a function of time for the different hadronic resonances discussed. The net gain rate is defined as
\begin{equation}
	\frac{1}{N_{\mathrm{CF}}}\left(\frac{d N_{\mathrm{gain}}}{dt} - \frac{d N_{\mathrm{loss}}}{dt}\right)
\end{equation}
where $N_{CF}$ is the resonance yield at hadronization (Cooper-Frye), $\frac{d N_{\mathrm{gain}}}{dt}$ is the number
of reactions that create a resonance and $\frac{d N_{\mathrm{loss}}}{dt}$ is the number of reactions that destroy a resonance, e.g. decay or inelastic scattering, per unit of time. The results presented in figure \ref{f3} clearly reflect the different lifetimes of the resonances, as the $\rho$ meson shows
its largest loss rate during early times, while the $\phi$ decays only slowly and therefore shows a slow loss. However there is an interesting comparison
between the $\Lambda^*$ and $\Sigma^*$. Even though there is a lifetime difference of a factor 2 between these states, their net gain rate appears to be very similar. This hints to the fact that the $\Sigma^*$ has a quite substantial regeneration rate at early times of the hadronic phase. Later we will see that this translates into specific different observable.  
 
\begin{figure}[t]	%       -----------------------------------------
\includegraphics[width=0.5\textwidth]{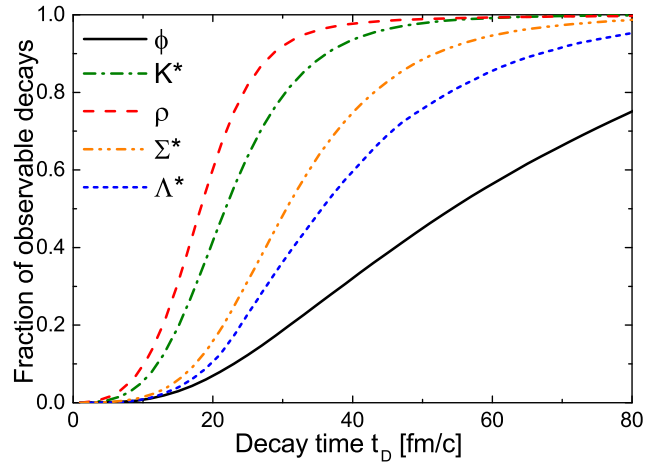}
\caption{[Color online] Fraction of observable resonances which have decayed until a time t. The results are for central collisions of Au+Au
at $\sqrt{s_{NN}}= 200 $ GeV.
}\label{f18}
\end{figure}		%       -----------------------------------------  

The integrated yield of the resonances normalized to their value at chemical freeze out, is shown in figure \ref{f4}, as a function of the computational time. Again we observe a clear hierarchy with respect to the lifetime of the resonance, with the exception of the $\Sigma^*$, which undergoes
significant regeneration in the time up to 20 fm/c.

The lifetime hierarchy can also be observed in figure \ref{f18} where we show the fraction of observable/reconstructable resonances which have decayed until a time $t_D$, as function of time. As expected, the $\phi$ meson decays at a very late time and the short lived resonances decay at an earlier time. As we have already discussed earlier the difference of the $\Sigma^*$ and  $\Lambda^*$ is smaller than expected due to the significant $\Sigma^*$ regeneration at early times. An important observation is that even though short lives resonances like the $\rho$ or $K^{*}$ are the first to decay, most of the observable resonances still decay at times $t> 15$ fm/c, i.e. a rather dilute hadronic system and hence may not carry much information on the dense system where they where born.

\begin{figure}[t]	%       -----------------------------------------
\includegraphics[width=0.5\textwidth]{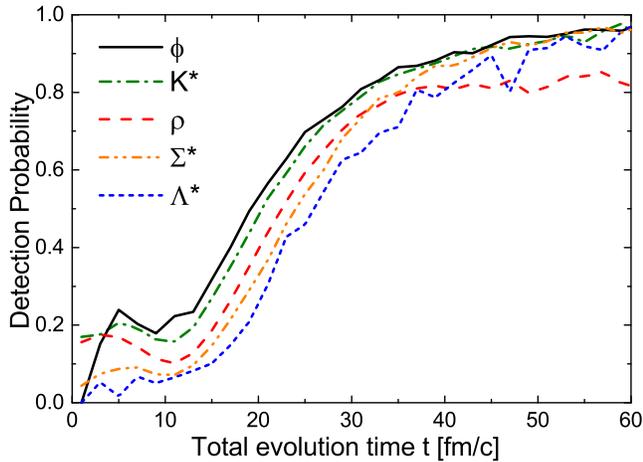}
\caption{[Color online] Detection probability of different hadronic resonances as function of the time of their decay. A resonance is detectable if none of their decay products rescatter in the hadronic phase. The results are for central collisions of Au+Au
at $\sqrt{s_{NN}}= 200 $ GeV.
}\label{f5}
\end{figure}

In order to use the resonances to understand the properties of the hadronic phase of a nuclear collision another aspect is crucial. Not only do these resonances have different lifetimes, but they also decay into different daughter particles which have to be identified in order to reconstruct the original resonance. As these daughter particles may also undergo rescattering they may be lost to reconstruction, thus also the information on the mother resonance would be lost. The probability of the daughter particle rescattering does depend on the time of the resonance decay and on the rescattering cross section of the daughter particles which is different for example for pions and kaons.
Within our model simulation we can investigate the rescattering of these daughter particles and define a detection probability, which gives the probability that the daughter particles of a given resonance did not rescatter subsequently in the hadronic medium.
This detection probability is shown as function of the total evolution time, in the center of mass frame of the collision, in figure \ref{f5}.
All considered resonances show a similar dependence on time of their detection probability. This is expected, as the rescattering of the daughter particles depends to first order on the density of possible scattering partners which strongly decreases with time. Furthermore the detection probability of the $\phi$ is slightly larger than that of the $\rho$ as the kaons from the $\phi$ have a smaller rescattering cross section than the pions from the $\rho$ decay.

\begin{figure}[t]	%       -----------------------------------------
\includegraphics[width=0.5\textwidth]{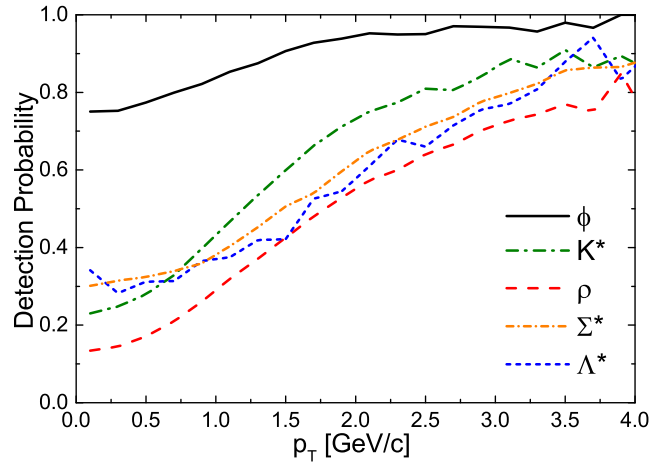}
\caption{[Color online] Detection probability of different hadronic resonances as function of their transverse momentum. A resonance is detectable if none of their decay products rescatter in the hadronic phase. The results are for central collisions of Au+Au
at $\sqrt{s_{NN}}= 200 $ GeV.
}\label{f17}
\end{figure}

As we have seen before, not all resonances decay at the same time and therefore the time integrated detection probability for the different resonances will depend also on their lifetime. In figure \ref{f17} we therefore show the time integrated detection probability as a function of the transverse momentum $p_{T}$ of the resonance. We observe two distinct features. First of all we see that there is a clear hierarchy of the detection probability of the resonances where the $\phi$ with its long lifetime and small hadronic cross section has the highest and the $\rho$ has the lowest probability of being detectable. Furthermore we also observe a strong dependence of the detection probability on the transverse momentum of the resonance. This can be understood because fast resonances leave the system more rapidly than slow ones and have therefore a lower chance that their decay products rescatter.

		%       -----------------------------------------  

\begin{figure}[t]	%       -----------------------------------------
\includegraphics[width=0.5\textwidth]{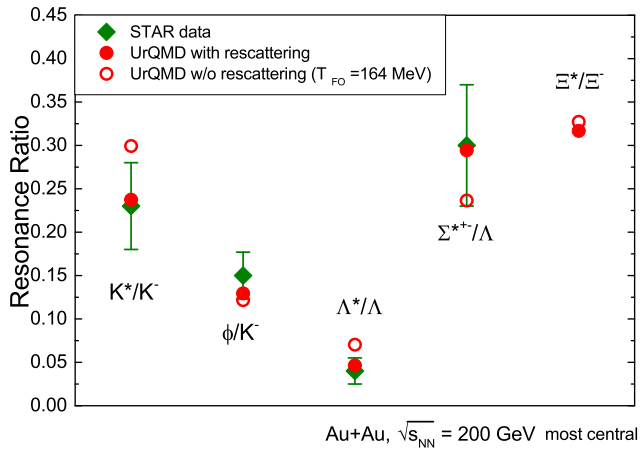}
\caption{[Color online] Ratios of detectable resonances to their ground state after the hadronic phase (red full circles) and before hadronic rescattering (red open circles), compared to data from the STAR experiment (green diamonds).
}\label{f6}
\end{figure}

Will the final state interactions lead to observable differences in resonance multiplicities and properties? To answer this question we calculate the ratio of the $p_{T}$ integrated multiplicity of detectable resonances and the corresponding multiplicity of their ground state hadron. We can do this at two times: either at the end of the hadronic phase by including all hadronic interactions or at the chemical freeze out by extracting this ratio after the hadron production on the Cooper-Frye hypersurface. The resulting ratios for central collisions of Au+Au at a beam energy of $\sqrt{s_{\mathrm{NN}}}= 200$ GeV, are depicted in figure \ref{f6}. Here we also compare the results with experimental data from the STAR experiment \cite{Adams:2004ep,Abelev:2008yz,Adams:2006yu}. We observe a very good agreement of our results with the experimental data in the case of the full model. When the final hadronic rescattering is neglected there is a deviation from the experimental results, especially for the $K^*$ and $\Lambda^*$ ratios, indicating that they are most sensitive to the rescattering phase.

\begin{figure}[t]	%       -----------------------------------------
\includegraphics[width=0.5\textwidth]{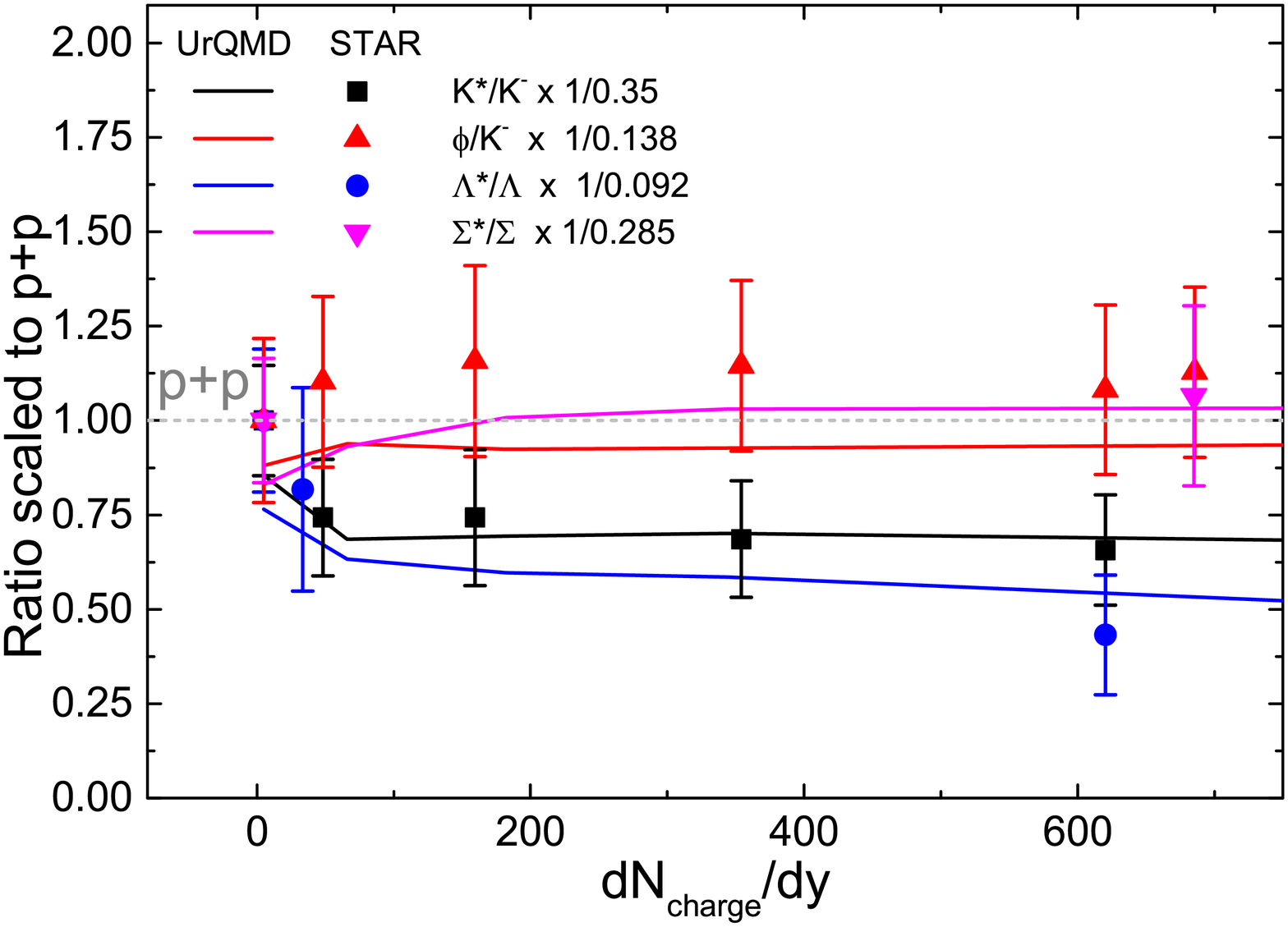}
\caption{[Color online] Ratios of detectable resonances to their ground state after the hadronic phase as function of centrality (lines) compared to experimental data (symbols). The results are for collisions of Au+Au at $\sqrt{s_{NN}}= 200 $ GeV. The quoted temperature $T_{FO}$ corresponds to the $\epsilon \approx 350$ MeV/fm$^3$ criterion.
}\label{f7}
\end{figure}

To confirm further the importance of hadronic rescatter, we show the centrality dependence of the aforementioned ratios in figure \ref{f7}. Here we see again that the $K^*$ and $\Lambda^*$ ratios decrease significantly from peripheral to central collisions. As expected, in the larger systems created in central collisions, the decay products have a larger probability to rescatter. Thus the resonance is effectively lost from reconstruction and the ratio is decreased.
An interesting result of the hadronic rescattering is that the $\phi/K$ and $\Sigma^*/\Lambda$ ratio actually increase slightly for central collisions while the other ratios decrease. For the $\phi$ this is a result of the long lifetime and small hadronic cross section as well as a small regeneration of $\phi$ mesons. For the $\Sigma^*$ this behavior is a result of its smaller mass, as compared to the $\Lambda^*$ and therefore of a higher regeneration rate at later times.

		%       -----------------------------------------  

		%       -----------------------------------------  

\section{Stable hadrons}

\begin{figure}[t]	%       -----------------------------------------
\includegraphics[width=0.5\textwidth]{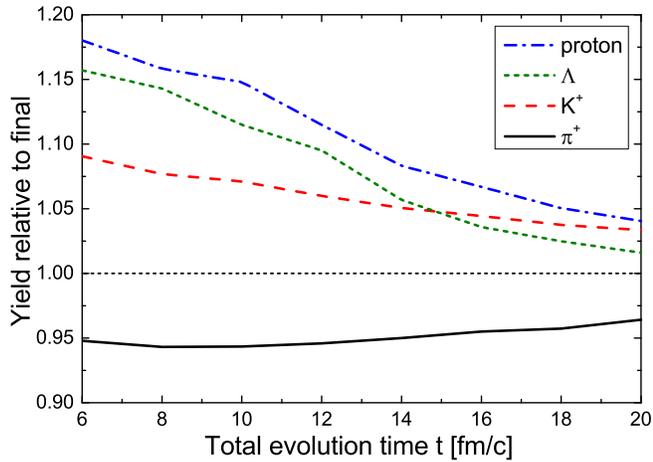}
\caption{[Color online] Time dependence of the stable hadron yields (including their resonance feed down), normalized to the final yields, during the hadronic phase. The results are for central collisions of Au+Au at $\sqrt{s_{NN}}= 200 $ GeV. 
}\label{f8}
\end{figure}

The observations on hadronic resonance yields strongly indicate the existence of a hadronic rescattering phase which has measurable impact on hadron properties. In the following we want to discuss whether we can also expect an impact of this rescattering phase on stable hadron yields and properties like spectra and collective flow. Here we can distinguish two types of hadronic scatterings. One is the (pseudo-)elastic scattering which changes only the momentum distribution of the present hadrons but not their abundances. Furthermore there may be inelastic scatterings which may change the flavor and hadronic abundances during the hadronic rescattering. The most relevant processes here are annihilation of baryons and anti-baryons, strangeness exchange reactions and multi-body decays of resonances essentially increasing the number of pions.
In figure \ref{f8} we show the time dependence of stable hadron yields (including contributions from resonance decays) as a function of time, for most central collisions of Au+Au at a beam energy of $\sqrt{s_{NN}}= 200$ GeV. Here the multiplicities are scaled to the final (observable) value to quantify the effect of the hadronic rescattering. We observe a very clear time dependence, confirming that the hadronic rescattering changes the chemical composition of the stable hadrons to times as large as 20 $fm/c$. We also observe characteristic differences for the different hadron species. For example the protons show the largest decrease in yield due to annihilation. The proton annihilation is stronger than that of $\Lambda$ hyperons due to the larger annihilation cross section. In fact we observe a clear hierarchy due to the decrease of the cross section with increasing strangeness content.

		%       -----------------------------------------  

\begin{figure}[t]	%       -----------------------------------------
\includegraphics[width=0.5\textwidth]{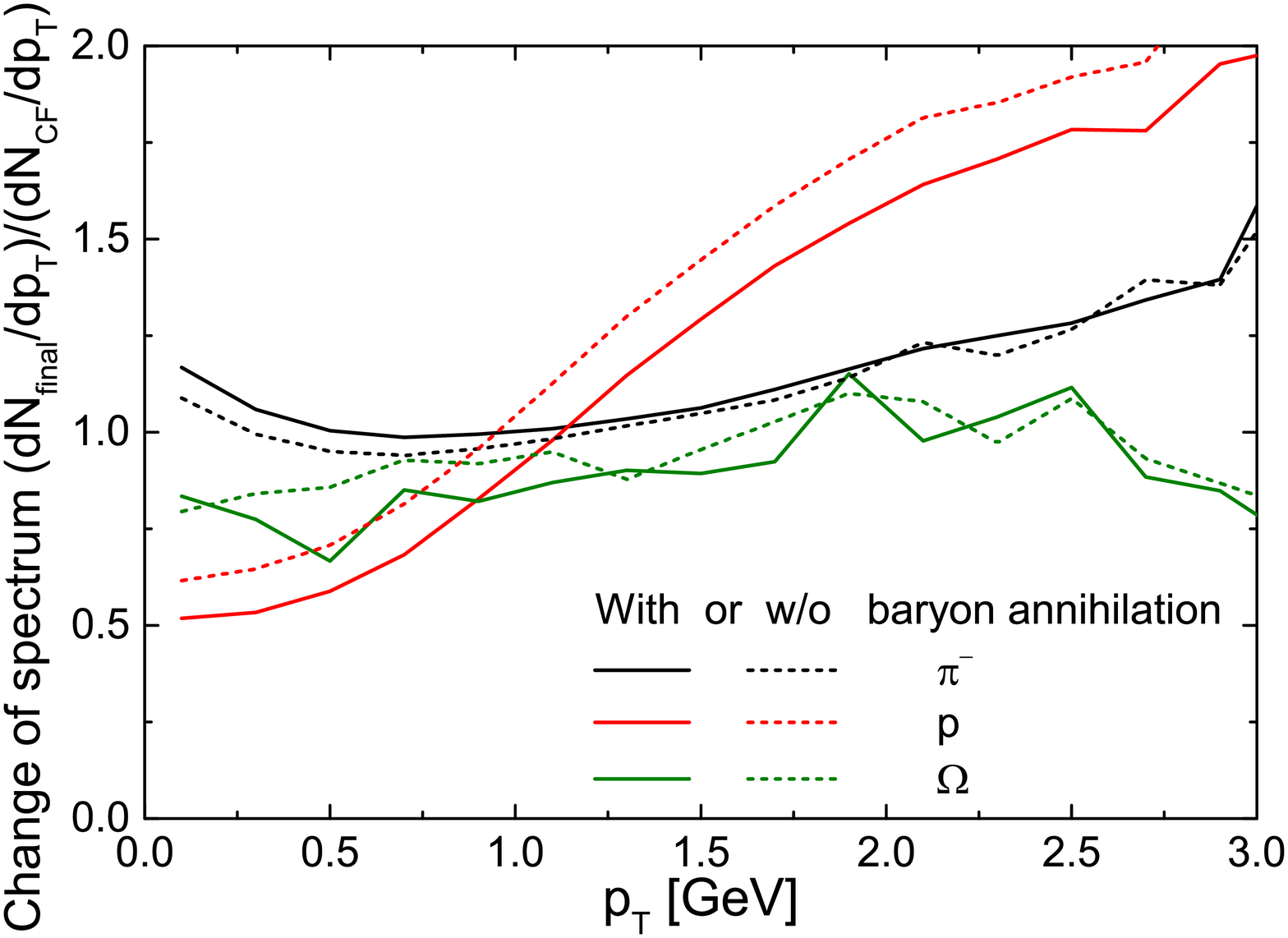}
\caption{[Color online] Change of the spectrum of stable hadrons due to the finals state rescattering. We compare two scenarios, one including the baryon+antibaryon annihilation and one excluding the annihilation reaction. The results are for central collisions of Au+Au
at $\sqrt{s_{NN}}= 200 $ GeV.
}\label{f9}
\end{figure}

The effect of hadronic rescattering on the final observed momentum spectra is shown in figure \ref{f9}. Here the change of the spectra is shown as the ratio of the final $p_T$ spectrum over the spectrum at chemical freeze out (Cooper-Frye). We compare three distinct hadrons, pions, protons and $\Omega$. Each species shows a very different characteristic change of the spectrum. The proton spectrum is mainly enhanced at large momenta and is depleted at low momenta, which reflects the generation of transverse collective velocity flow which has been observed in these reactions. This change is mainly caused by elastic and pseudo-elastic scatterings, while baryon antibaryon annihilation decreases the spectrum mostly independent of $p_T$. The pions also obtain an additional radial flow, however also the low $p_T$ part of the pion spectrum is enhanced. This is due to contributions from resonance decays which populate mainly the low transverse momenta. The $\Omega$ baryon receives the smallest change in the spectrum due to its small hadronic rescattering cross section. It is little affected from the transverse flow. That the average value of the ratio is below one is likely a consequence of the strangeness exchange reaction and longitudinal expansion.

\section{Annihilation}

An important and much discussed contribution to the final state hadronic rescattering is the baryon+anti-baryon annihilation reaction. In this reaction the baryons, e.g. proton and anti-proton, annihilate to form multiple mesons. For the case of nucleons and anti-nucleon on average about 5 pions are created. Since most transport models (as UrQMD) do not include multi particle scatterings, the inverse reaction, 5 pions creating a baryon + anti-baryon pair, is not taken into account \cite{Rapp:2000gy,Greiner:2000tu,Cassing:2001ds,Satarov:2013wga}. This violates detailed balance but in an expanding system which hadronizes at an energy density of $350 MeV/fm^3$ this reaction, which decreases with $\rho^5$, is rare \cite{BraunMunzinger:2003zz}. This has also been shown in studies where the detailed balance for this reaction has been explicitly included and the resulting baryon and anti-baryon loss in the hadronic phase is comparable to our scenario \cite{Pan:2012ne}. Nevertheless it is worthwhile to further pursue the implementation of a hadronic rescattering phase that includes such multi-particle scattering in order to minimize the systematic errors on the model side.

		%       -----------------------------------------  

		%       -----------------------------------------  

\begin{figure}[t]	%       -----------------------------------------
\includegraphics[width=0.5\textwidth]{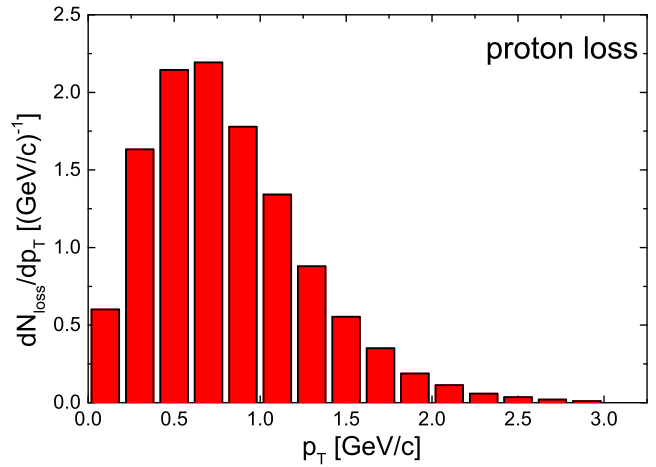}
\caption{[Color online]Proton loss rate (by annihilation) as a function of the transverse momentum of the proton in the center of mass frame of the heavy ion collision. The results are for central collisions of Au+Au
at $\sqrt{s_{NN}}= 200 $ GeV.
}\label{f10}
\end{figure}

\begin{figure}[t]	%       -----------------------------------------
\includegraphics[width=0.5\textwidth]{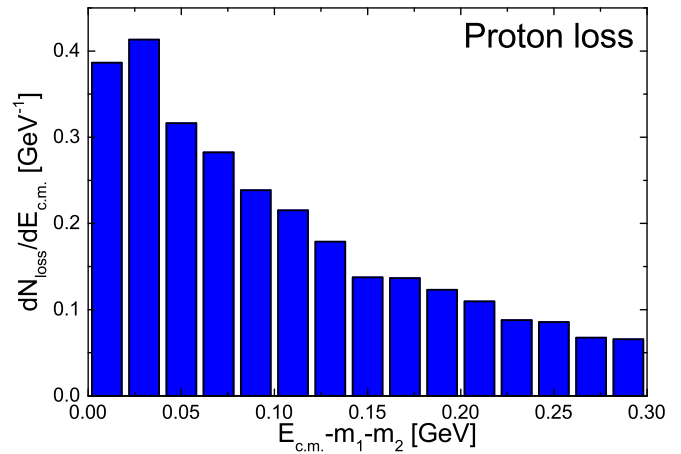}
\caption{[Color online] Proton loss rate (by annihilation) as a function of the invariant mass (minus the rest masses of the annihilating baryons $m_1$ and $m_2$) of the baryon+antibaryon system in the center of mass frame of the annihilation reaction. The results are for central collisions of Au+Au at $\sqrt{s_{NN}}= 200 $ GeV.
}\label{f11}
\end{figure}

In order to understand why the relative proton loss is mainly independent of transverse momentum, even though the annihilation cross section is largest at small relative momenta, we will investigate the proton loss as function of $p_T$ (in figure \ref{f10}) and as function of the relative energy (figure \ref{f11}). We can clearly see that the loss rate has a maximum for a transverse momentum of roughly 600-700 MeV. This maximum also corresponds to the maximum in the proton transverse momentum distribution, indicating that the annihilation probability depends mainly on the phase space density. Consequently the relative loss of protons is independent of $p_T$. On the other hand we also clearly see that the loss of protons is maximal for small relative energies, as expected from the energy dependent annihilation cross section.

		%       -----------------------------------------  

		%       -----------------------------------------  

\begin{figure}[t]	%       -----------------------------------------
\includegraphics[width=0.5\textwidth]{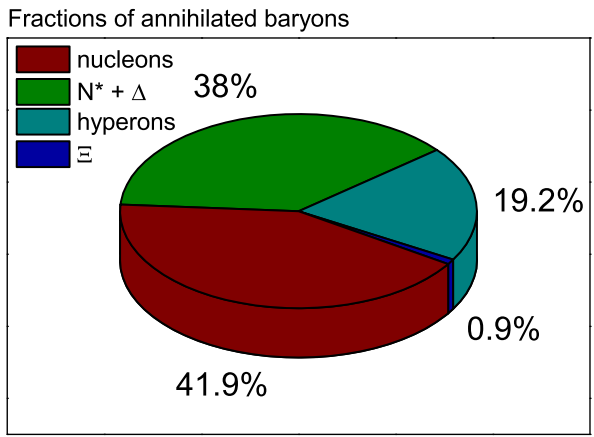}
\caption{[Color online] Relative fraction of different baryonic species lost due to baryon+antibaryon annihilation in the hadronic phase. Only 42$\%$ of all annihilated baryons are actual nucleons. The rest is composed of excited nucleonic states as well as hyperons and their resonance states. As a result less than 25 percent of all annihilations actually takes place between two nucleons in their ground state. The results are for central collisions of Au+Au at $\sqrt{s_{NN}}= 200 $ GeV.
}\label{f12}
\end{figure}

Finally we want to address an important aspect of baryon annihilation in the hadronic phase which has, so far, been neglected in all studies. For simplicity one usually argues about the annihilation of ground state nucleons, and possibly hyperons, but generally one neglects that a significant portion of baryons consists of excited resonance states with significant larger masses. For example at LHC and RHIC at chemical freeze out almost half of all baryon exist in form a baryonic resonance. In our approach the absolute value and the $\sqrt{s}$ dependence of the annihilation cross section for the resonances is assumed to be identical to that of the ground state with the same quantum numbers. In figure \ref{f12} we show the percentage of the different annihilation channels which contribute to the annihilation. One can clearly see that less than half of all annihilated baryons are actually in their ground state. This also means that in less than 25$\%$ of all annihilation reactions two ground state baryons annihilate. The majority of these reactions will have a larger invariant mass and therefore produce on average more than 5 pions and eventually even heavier mesons like kaons. Thus the back reaction of these annihilation reactions will be even more suppressed by the phase space density and also the effective implementation of all possible processes is much more involved.

		%       -----------------------------------------  

\section{flow}

In this last section we want to discuss the effect of the hadronic rescattering on the identified particle elliptic flow measurements. As we have seen in the previous section, the (pseudo)-elastic scattering of hadrons leads to a significant change in the $p_T$ distribution, similar to an increase of radial flow. In the following we will discuss
how the elliptic flow of identified particles, defined as
\begin{equation}
v_2=\left\langle\frac{p_{x}^{2}-p_{y}^{2}}{p_{x}^{2}+p_{y}^{2}}\right\rangle
\end{equation}
is modified during the hadronic phase.

\begin{figure}[t]	%       -----------------------------------------
\includegraphics[width=0.5\textwidth]{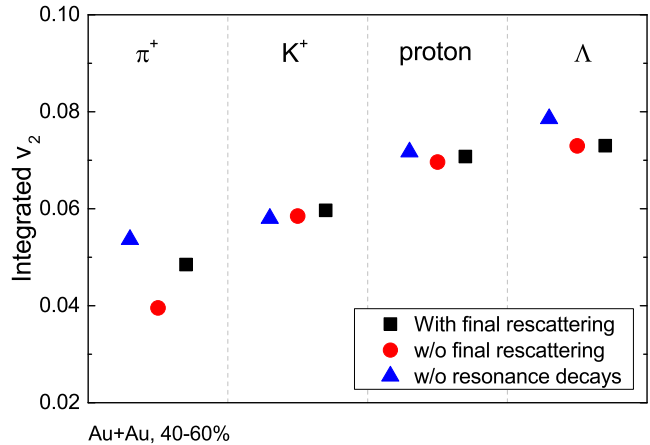}
\caption{[Color online] The $p_T$-integrated elliptic flow $v_2$ for different hadronic particle species. We compare the elliptic flow of the hadrons at chemical freeze out (Cooper-Frye) excluding (blue triangles) and including freed down from the resonances (red circles) with the value after the hadronic phase (black squares). The results are for central collisions of Au+Au at $\sqrt{s_{NN}}= 200 $ GeV.
}\label{f13}
\end{figure}

Here the average runs over all particles in an event and then also over all considered events. For simplicity we will calculate the elliptic flow with respect to the reaction plane of the collision.\\
In figure \ref{f13} we present the results for the $p_T$ integrated elliptic flow for several identified particles, in $40 - 60 \%$ central collisions of Au+Au at a beam energy of $\sqrt{s_{\mathrm{NN}}}=200 $~GeV.
Here we compare the $p_T$ integrated $v_2$ for different times during the evolution.
\begin{enumerate}
\item Blue triangles: at chemical freeze out (cooper-Frye). In this case we only calculate the elliptic flow from primordially produced hadrons, i.e. we neglect the contribution from feed down due to resonance decays during the hadronic phase.
\item Red circles: at freeze out (Cooper-Frye), this time including the contributions from resonance decays.
\item Black squares: At the end of the hadronic phase. This is the $v_2$ which can be compared to experimental data.
\end{enumerate}

		%       ----------------------------------------- 

\begin{figure*}[t]	%       -----------------------------------------
\includegraphics[width=\textwidth]{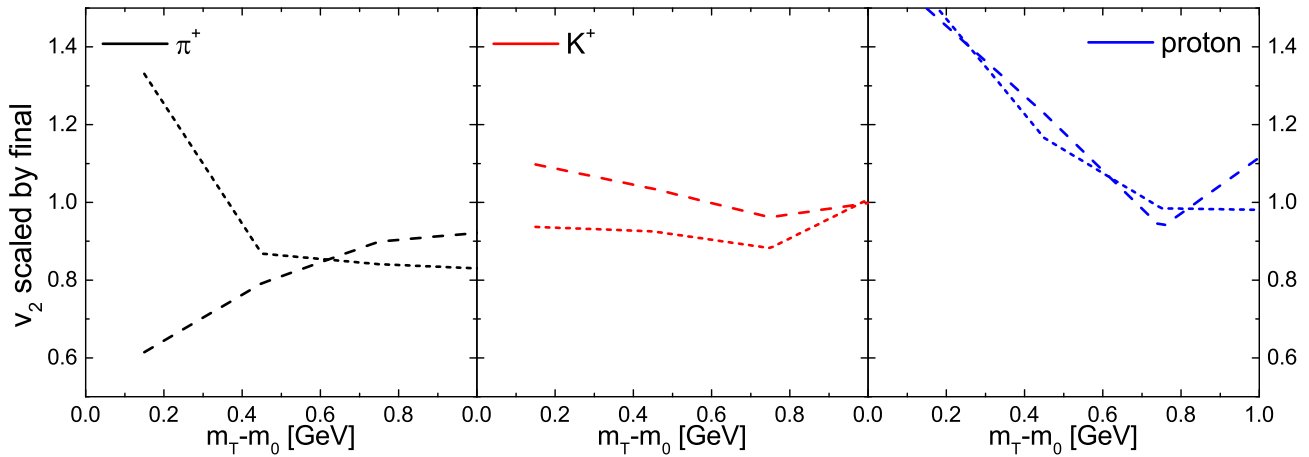}
\caption{[Color online] Relative change of the elliptic flow $v_2$ for different particle species as a function of the transverse mass $m_T$. We divided the elliptic flow of the hadrons at the chemical freeze out (Cooper-Frye) excluding (short dashed lines) and including freed down from the resonances (dashed lines) with the value after the hadronic phase. The results are for mid-central (40-60$\%$) collisions of Au+Au
at $\sqrt{s_{NN}}= 200 $ GeV.}\label{f14}
\end{figure*}		%       ----------------------------------------- 

We observe a strong dependence of the pion integrated elliptic flow on the different contributions. As expected, the resonance contributions decrease the elliptic flow, while the hadronic rescattering increases it again. The kaons and protons appear less sensitive on the different effects while the $\Lambda$ is mainly sensitive in resonance contributions, which are significant for the $\Lambda$.

Finally we want to study the $m_T$ dependence of the elliptic flow.  We show the ratio of the $m_T$ differential elliptic flow, at different stages of the systems evolution, for pions, kaons and protons in figure \ref{f14} with respect to the final observed elliptic flow. Here again the change of the elliptic flow during the hadronic phase is small for the kaons a consequence of the small hadronic scattering cross section. For protons the hadronic phase lowers $v_2$ for low and intermediate transverse energies. The pions suffer an even larger change of $v_2$ but in opposite direction, we observe an increase of $v_2$ during the hadronic phase which is about constant for all transverse energies.  This also means that the increase of the integrated $v_2$ is due to the increase in low $p_T$ pions (see figure \ref{f9}) from the hadronic rescattering, and not from a decrease in the $p_T$ dependent $v_2$.

\section{Summary}
We have presented a comprehensive summary of the dynamics in the hadronic re-scattering phase of the UrQMD transport model. The hadronic re-scattering has a substantial ($\sim 30\%$) impact on most hadronic observables. In particular the calculated resonance yields which reproduce well the experimental data do not only verify the existence of the hadronic phase experimentally but are also a sensitive probe of the interactions taking place in the hadronic phase. Resonances like the $K^*$, with a rather short lifetime, show a significant reduction in the observed yield. In particular at low $p_T$ the measurable resonance yields are reduced by up to 80$\%$. The $\phi$ production rate stays essentially constant throughout the hadronic phase due to its long lifetime and its small hadronic cross section. Furthermore the spectra and yields of stable hadrons are changed significantly, e.g. the proton number decreases by about 20$\%$ due to the annihilation of nucleons and excited baryonic states with their anti-particles.
We also observe significant ($\sim 20\%$) modifications for the elliptic flow $v_2$ during the hadronic phase.
These observations show that the experimental results cannot assert, or be directly compared to, the properties of the system at a conjectured hadronization or chemical freeze-out point.

\section*{Acknowledgments}
We thank V. Vovchenko for helpful comments.
This work was supported by GSI. The computational resources were provided by the LOEWE Frankfurt Center for Scientific Computing (LOEWE-CSC).

%%%%%%%%%%%%%%%%%%%%%%%%%%%%%%%%%%%%%%%%%%%%%%%%%%%%%%%%%%%%%%%%%%%%%%%%%%%%%%%
%%%%%%%%%%%%%%%%%%%%%%%%%%%%%%%%%%%%%%%%%%%%%%%%%%%%%%%%%%%%%%%%%%%%%%%%%%%%%%% 

\end{document}